\documentclass[aps,prb,reprint,superscriptaddress]{revtex4-1}

\newcommand{\euya}{\mbox{Eu$_{0.8}$Y$_{0.2}$MnO$_3$}}

\newcommand{\euyc}{\mbox{Eu$_{\nicefrac{3}{4}}$Y$_{\nicefrac{1}{4}}$MnO$_3$}}
\newcommand{\euyx}{\mbox{Eu$_{1-x}$Y$_{x}$MnO$_3$}}
\newcommand{\rmo}{\mbox{{\it R}MnO$_3$}}
\newcommand{\tbmo}{\mbox{TbMnO$_3$}}

\newcommand{\vecc}[1]{\mathbf{#1}}

\newcommand{\pisi}{\mbox{$\pi$-$\sigma'$}}
\newcommand{\pipi}{\mbox{$\pi$-$\pi'$}}

\newcommand{\sfz}{\mbox{$^7F_0$}}
\newcommand{\sfo}{\mbox{$^7F_1$}}

\usepackage{hyperref}
\usepackage{amssymb,amsmath,units,bm} 
\usepackage{graphicx}
\usepackage{xcolor}
\usepackage[english]{babel}

\begin{document}


\title{Long range antiferromagnetic order of formally non-magnetic
  Eu$^{3+}$ Van Vleck ions observed in multiferroic \euyx}


\author{A. Skaugen}
\email[]{arvid.skaugen@desy.de}
\affiliation{Deutsches Elektronen-Synchrotron DESY, Notkestra{\ss}e 85, D-22607 Hamburg, Germany}
\author{E. Schierle}
\affiliation{Helmholtz-Zentrum Berlin f\"ur
Materialien und Energie, Albert-Einstein-Str. 15, D-12489 Berlin, Germany}
\author{G. van der Laan}
\affiliation{Diamond Light Source, Chilton, Didcot OX11 0DE, United Kingdom}
\author{D. K. Shukla}
\affiliation{Deutsches Elektronen-Synchrotron DESY, Notkestra{\ss}e 85, D-22607 Hamburg, Germany}
\affiliation{UGC DAE Consortium for Scientific Research, Khandwa Road, Indore 01, India}
\author{H. C. Walker}
\affiliation{Deutsches Elektronen-Synchrotron DESY, Notkestra{\ss}e 85, D-22607 Hamburg, Germany}
\affiliation{ISIS, Rutherford Appleton Laboratory, Chilton, Didcot OX11 0QX, United Kingdom}
\author{E. Weschke}
\affiliation{Helmholtz-Zentrum Berlin f\"ur
Materialien und Energie, Albert-Einstein-Str. 15, D-12489 Berlin, Germany}
\author{J. Strempfer}
\affiliation{Deutsches Elektronen-Synchrotron DESY, Notkestra{\ss}e 85, D-22607 Hamburg, Germany}

\date{\today}

\begin{abstract}

  We report on resonant magnetic x-ray scattering and absorption
  spectroscopy studies of exchange-coupled antiferromagnetic ordering
  of Eu$^{3+}$ magnetic moments in multiferroic \euyx\ in the absence
  of an external magnetic field.  The observed resonant spectrum is
  characteristic of a magnetically ordered \sfo\ state that mirrors
  the Mn magnetic ordering, due to exchange coupling between the Eu
  $4f$ and Mn $3d$ spins.  This is the first observation of long range
  magnetic order generated by exchange coupling of magnetic moments of
  formally non-magnetic Van Vleck ions, which is a step further
  towards the realization of exotic phases induced by exchange
  coupling in systems entirely composed of non-magnetic ions.
  
\end{abstract}


\maketitle



The interplay between local spin and orbital magnetic moments is an
important factor in a large variety of magnetic ordering phenomena
exploited in present day applications.  In special cases, even with
magnetic moments present, a system can form a non-magnetic singlet
ground state.  Prominent examples are rare earth and transition metal
ions with the $f$ or $d$ shell missing one electron for half filling.
Here the orbital and spin moments can cancel out generating a $J=0$
ground state.  However, having only a small energy spacing between the
ground state and the first magnetic triplet state, magnetism can in
such systems be generated by symmetry breaking external stimuli like
magnetic or electric fields.  In the case of magnetic stimulus this is
known as Van Vleck magnetism and offers fascinating possiblities for
new applications, for example for a magnetic sensor that is itself
non-magnetic.  From a more fundamental point of view such systems are
candidates for a variety of novel states of matter characterized by
hidden order~\cite{Ada99}, Bose-Einstein condensation or quantum phase
transitions~\cite{Rue03,Zap06}.

A paramount example of a formally non-magnetic ion being susceptible
to external magnetic fields is the Eu$^{3+}$ ion with $S=3$ and $L=3$,
having a $J=0$ non-magnetic ground state.  Van Vleck magnetism has
long been known to contribute to the paramagnetic moment of
Eu$^{3+}$~\cite{Fra32,vanVleck68,Tagirov2002,Tak10}.  For this ion, the symmetry
breaking by an external magnetic field mixes the \sfo\ state into the
ground state, yielding a finite magnetic moment.  A more recent
example of this mechanism is the observation of x-ray magnetic
circular dichroism (XMCD) in EuN under an applied magnetic field of 5
T, which is explained by magnetic field induced admixture of \sfo\
into the \sfz\ ground state~\cite{Ruc11}.  The possibility of spin
ordering without external electric or magnetic stimulus in the case of
a vanishing total magnetic moment has been discussed theoretically,
setting up the possibility of an unconventional phase transition in
which the spin correlation length diverges but there is little or no
change in the magnetic properties~\cite{Joh05}.

While these previous studies revealed the presence of Van Vleck
magnetic moments, experimental proof of intrinsic long-range magnetic
order of Van Vleck ions is missing.  The perovskite-structure rare
earth (RE) manganites \rmo\ are well-suited candidates to show such a
mechanism.  These compounds have attracted much attention due to their
strong magnetoelectric (ME) effect and the possibility to control
electric (magnetic) order by magnetic (electric)
fields~\cite{Fie05,Kim03,Got04,Kim05}.  These multiferroic properties
are largely related to the magnetic order of the Mn $3d$ magnetic
moments~\cite{Mos06,Moc09}.  However, RE magnetic ordering has been
shown to play a decisive role in the multiferroicity of these
compounds in the past years~\cite{Fey09,Fey10,Ska14,Sch10,Wal11}.  In
systematic studies of such complex compounds, a standard way to
disentangle various magnetic contributions is the comparison with a
compound involving a non-magnetic RE.  In multiferroics, the \euyx\
series of compounds is a prominent example~\cite{Hem07,Yam07,Moc09}.
The $J=0$ ground state of Eu$^{3+}$ is non-magnetic and has spherical
symmetry, so the crystal field splitting is expected to be small.
However, exchange coupling between Mn $3d$ and RE $4f$ states is an
interaction between spins and does not involve the orbital moment $L$,
and may hence induce a Van Vleck $J=1$ magnetic moment in Eu$^{3+}$.
As a consequence, Eu$^{3+}$ might not be anticipated as a completely
nonmagnetic reference ion in exchange coupled materials, but displays
a perfect candidature for magnetic order originating from Van Vleck
magnetism.

In this Paper, we demonstrate long range complex antiferromagnetic
order of Eu$^{3+}$ ions without invoking an external electromagnetic
field or macroscopic magnetization.  We used resonant elastic x-ray
scattering (REXS) to study the magnetic reflections from \euya.  REXS
is particularly suited for this purpose by virtue of element
specificity, high sensitivity to detect even weak antiferromagnetic
ordering of Eu $4f$ moments, and spectroscopic information for
identifying the $J$ state involved in the ordering.

The soft x-ray experiments were carried out using the XUV
diffractometer and the High-field diffractometer at beamline UE46-PGM1
at the \mbox{BESSY II} storage ring~\cite{Eng01,Fin13}, while the hard
x-ray data were taken at beamline P09 at the \mbox{PETRA III} storage
ring at DESY~\cite{Str13}.  The scattering experiments at both
facilities were carried out in horizontal scattering geometry from the
polished $b$ surface of the sample~\cite{Iva06}, at \mbox{PETRA III}
with $a$ perpendicular to the scattering plane, and at \mbox{BESSY II}
with $c$ perpendicular to the scattering plane, unless otherwise
noted.  For cooling the sample, a 14 T cryomagnet and a Displex
cryostat were used at P09, and a continuous flow LHe cryostat was used
at UE46-PGM1.  For the hard x-ray experiments a Cu 220 polarization
analyzer was mounted behind the sample to separate the \pipi and \pisi
channels and to suppress the fluorescence background.  Here $\pi$
($\pi'$) and $\sigma$ ($\sigma'$) denote polarization directions of
the incoming (outgoing) beam parallel and perpendicular to the
scattering plane, respectively~\cite{Hil96}.  At \mbox{BESSY II} the
incident polarization was varied, and the total scattered beam was
detected.

The Mn magnetic structure of \euyc\ below $T_C \approx 28\ \mathrm{K}$
has been shown to consist of an A type $ab$ plane cycloid with a wave
vector $\bm{\tau} = \nicefrac{1}{4} \mathbf{b}^*$ in the $Pbnm$
orthorhombic unit cell, upon which an F type $c$ axis sinusoid is
superposed through the Dzyaloshinskii-Moriya (DM)
interaction~\cite{Jan11}.  The same mechanism is also expected to
induce a G type $ab$ plane cycloid and a C type $c$ axis sinusoid.
\euya\ was investigated at low temperature in the ferroelectric phase.
Measuring the REXS intensity at the Mn $K$ edge, magnetic A, F, G and
C type reflections were observed.  For the C type reflection at the Mn
$K$ edge, a very weak signal was found in the \pisi channel only,
which indicates that the Mn moments are oriented along $c$.  In
contrast, at the Mn $L_{2,3}$ edges only the F type reflection is
accessible within the limited size of the Ewald sphere at this photon
energy.

Remarkably, intense resonant F and C type reflections could be
observed at the Eu $M_{4,5}$ edges as well.
\begin{figure}[t]
  \includegraphics[width=\linewidth]{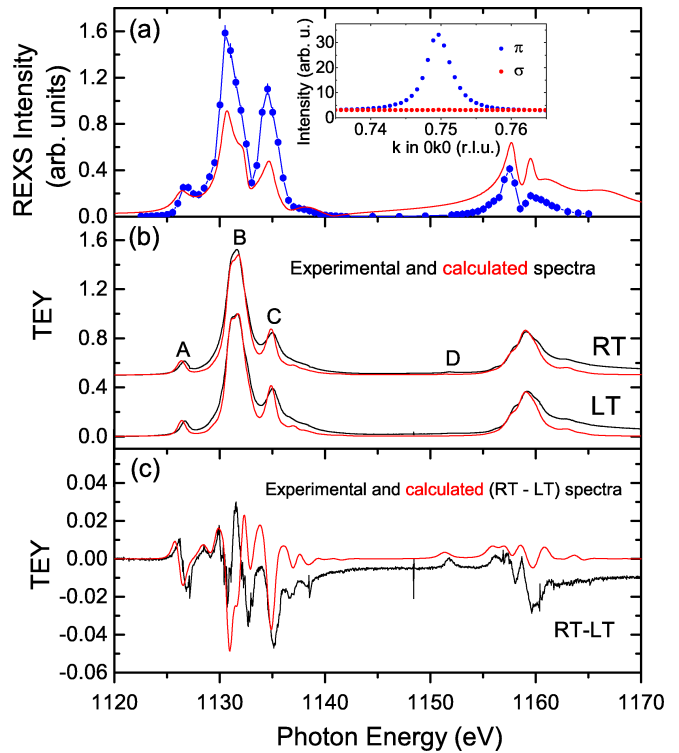}%
  \caption{(a) (Color online) Absorption corrected REXS spectrum (blue
    (dark grey) curve) of \mbox{(0 1-$\tau$ 0)} across the Eu $M_5$
    and $M_4$ edges at $T$ = 10 K, and lineshape calculation (red
    (light grey) curve) based on XMCD spectrum from Ref.
    \onlinecite{Ruc11}.  The inset shows reciprocal space scans of the
    reflection for $\pi$ and $\sigma$ incident polarization. (b)
    Experimental TEY spectra and multiplet calculations at $T$ = 296 K
    (RT) and $T$ = 120 K (LT).  (c) Measured and calculated difference
    between high and low temperature spectra.
    \label{fig:energy}}
\end{figure}
The resonant behavior at the Eu$^{3+}$ $M_{4,5}$ absorption edges is
demonstrated in Fig. 1.  The inset of Fig.~\ref{fig:energy}(a) shows a
reciprocal space scan over the C type \mbox{(0 1-$\tau$ 0)} reflection
at $T = 10~\mathrm{K}$ with $\pi$ and $\sigma$ incident polarization,
respectively, and the photon energy tuned to 1127 eV, close to the Eu
$M_5$ absorption edge.  The absence of intensity for $\sigma$ incident
light indicates that this reflection is caused by magnetic scattering
from the formally non-magnetic Eu$^{3+}$ ions with the corresponding
moments pointing along the $c$ direction, i.e. moments parallel to the
Mn moments as observed at the Mn $K$ edge resonance.  Resonant
enhancement at formally non-magnetic anions has been observed in
earlier studies and mainly been attributed to transferred moments in
hybrid orbitals~\cite{Mannix2001,Bea10,Par11}.  In contrast to all
former observations however, the Van Vleck ion Eu$^{3+}$ has the
possibility to create a magnetic moment in the core-like $4f$ shell by
populating the magnetic \sfo\ excited state.  The photon energy
dependence of the C type magnetic reflection is shown in
Fig.~\ref{fig:energy}(a), and can be readily explained by considering
the resonant magneto-optical parameters connected with a \sfo\ state.
The expected lineshape is calculated on the basis of experimental XMCD
data~\cite{Ruc11} of paramagnetic Eu$^{3+}$ and invoking a
Kramers-Kronig transform~\cite{Hav08}; the result is shown as red
curve in Fig.~\ref{fig:energy}(a).  Comparison to the measured REXS
lineshape (blue curve) yields a good match, apart from a relative
difference in intensities between the $M_4$ and $M_5$ edges and a
small offset in photon energy.  This result shows that the peak
observed at the Eu $M_{4,5}$ edges is indeed of magnetic origin caused
by a populated \sfo\ state.  To clarify the mechanism populating the
\sfo\ state we performed x-ray absorption spectroscopy (XAS) at
different temperatures by measuring the total electron yield (TEY)
(Fig.~\ref{fig:energy}(b)).  We then compared the TEY data to single
ion atomic multiplet calculations of the relevant electronic
states~\cite{Tho85}.

Fig.~\ref{fig:energy}(b) shows the x-ray absorption spectra across
the Eu $M_{4,5}$ edges at two different temperatures, 296 K (RT) and
120 K (LT).  The spectra are normalized such that the integrated
intensity of the difference spectrum in Fig.~\ref{fig:energy}(c) is
perceived to be as small as possible.  Due to the sample being
ferroelectric and thus insulating at low temperatures, 120 K was the
lowest temperature we were able to measure TEY spectra without
encountering charge buildup on the surface.

\begin{figure}[t]
  \includegraphics[width=\linewidth]{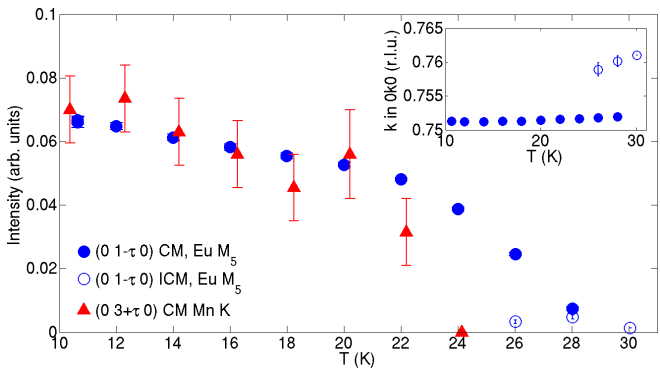}%
  \caption{(Color online) Temperature dependence of the integrated
    intensity of the C type \mbox{(0 1-$\tau$ 0)} reflection at the Eu
    $M_5$ resonance (blue circles). Red triangles show the $T$
    dependence of the C type \mbox{(0 3+$\tau$ 0)} reflection at the
    Mn $K$ edge. The data have been scaled to fit onto the same plot.
    The inset shows the peak positions in $q$-space over the same
    temperature range.\label{fig:tdep}}
\end{figure}

These spectra are compared to multiplet calculations $4f^6 \to 3d^9
4f^7$, performed assuming Boltzmann population of the different $J$
levels.  The calculations result in 26.9\% and 1.7\% \sfo\ population
at RT and LT, respectively, as expected from the thermal population
($E_{J=1} - E_{J=0} = 53 \mathrm{meV}$).  There are some striking
differences between the RT and LT spectra seen in the experiment which
are fully confirmed by the calculation: The peaks A, C and the
low-energy shoulder of the main peak B are all lower for RT, and a
small additional peak D appears for RT.  Altogether, there is an
excellent agreement between experiment and calculation for the (RT-LT)
difference spectra shown in Fig.~\ref{fig:energy}(c), taking into
account some small peak shifts, asymmetric line shapes, and continuum
background not included in the calculation.  Furthermore, the
magnitudes of the experimental and calculated differences are in good
agreement with each other.  The maximum difference is $\sim$ 5 \%.
The magnitude of the difference spectrum scales with the percentage of
higher $J$ population.  Thus thermal population of $J \neq 0$ states
can not explain the resonant signal at 10 K.  The analysis puts an
upper limit of $\sim$ 10 \% to a temperature independent \sfo\
contribution, i.e. from hybridization and crystal field
effects.

The negligible influence of the temperature on the \sfo\ population is
also reflected in the observed temperature dependent peak intensity.
The temperature dependence of the C type reflection (0 1-$\tau$ 0) was
measured both at the Mn $K$ edge and at the Eu $M_5$ edge, and is
shown in Fig.~\ref{fig:tdep}.  Apart from smaller error bars for the
much stronger signal at the Eu $M_5$ absorption edge,we observe a
perfect match of the temperature dependent peak intensities,
disappearing at the transition into the paraelectric phase.  This
behavior is in accordance with the proposed magnetic structure; the
behavior of the Eu$^{3+}$ ions, in particular, reflects that of the
corresponding order parameter.  This demonstrates a strong coupling
between the Mn and the Eu$^{3+}$ magnetic order capable of breaking
the symmetry of the $4f$ wave function.

The stronger signal at the Eu $M_5$ resonance reveals a second
incommensurate structure close to the transition into the paraelectric
phase, with a temperature-dependent peak position as shown in the
inset of Fig.~\ref{fig:tdep}.  This is also found in respective data
recorded at the Eu L$_{2,3}$ edges, where in addition, A, C, F and G
type reflections were also observed.  While C and F type order are
expected to polarize the induced Eu$^{3+}$ moments in this compound,
the observation of A and G type Eu order is more surprising, since Eu
in the $Pbnm$ space group sits at the crystallographic mirror
perpendicular to the $c$ axis. This means that the exchange field from
the A and G type Mn order, both being antiferromagnetically ordered
along c, would cancel out at the Eu site, and A and G type order of
the \sfo\ moments can in theory not be induced. However, similar
observations have also been made in earlier studies on \tbmo, where
the Tb moments mirror the A type Mn magnetic structure in the
collinear phase~\cite{Mannix2007,For08}.  Possible explanations,
already discussed in the above mentioned references, are either a
non-magnetic origin of these reflections, or ionic displacements of
the Eu ions from their ideal positions which could either lift the
strict extinction rules for the A and G type reflections or even break
the crystal symmetry such that A and G type order can be induced at
the Eu sites. As known from other \rmo\ compounds, varying amounts of
frustration are introduced to the crystal structure by substituting in
rare earth ions with different ionic radii. It is therefore reasonable
to expect that the $Pbnm$ symmetry in the current sample might be
broken such that the Eu is no longer located exactly on a
crystallographic mirror. Our data are fully consistent with such a
scenario, making a non-magnetic origin rather less likely. However, if
ionic displacement of the Eu ions is the correct explanation, our data
do not allow the identification of the type of ionic displacement nor
the driving mechanism behind it.

\begin{figure}[t]
  \includegraphics[width=\linewidth]{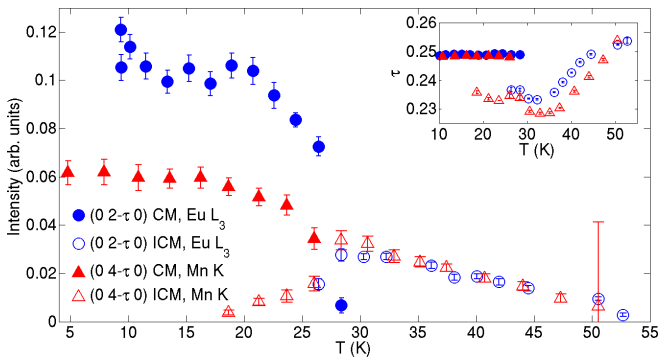}%
  \caption{(Color online) Temperature dependence of the integrated
    intensity of the F type \mbox{(0 2-$\tau$ 0)} reflection at the Eu
    $L_3$ absorption edge (blue circles).  The inset shows the peak
    positions in $q$-space in the same temperature range. Red
    triangles show the temperature dependence of the F type \mbox{(0
      4-$\tau$ 0)} reflection at the Mn $K$ edge.  The data have been
    scaled to fit onto the same plot.\label{fig:tdepF}}
\end{figure}

Since the resonant signal at the Eu $L_{2,3}$ edges does not directly
probe the $4f$ states, but rather the $5d$ electrons, we do not gain a
noteworthy signal enhancement compared to the Mn $K$ edge resonance
for these reflections.  Nonetheless, it allows the comparison of all
the relevant reflections from Eu$^{3+}$ with those observed at the Mn
$K$ edge.  Fig.~\ref{fig:tdepF} shows the integrated intensities of
the F type reflections (0 2-$\tau$ 0) and (0 4-$\tau$ 0) as functions
of temperature at the Eu $L_3$ and Mn $K$ absorption edges,
respectively.  The commensurate (CM) and incommensurate (ICM) phases
are recognized in both temperature spectra, and match the phase
transitions observed elsewhere~\cite{Hem07,Yam07,Iva06}.  The only
major difference in the data is the relative difference in intensity
between the CM and the ICM reflections, being slightly larger at the
Eu $L_3$ edge.

We thus observe a common behaviour of the magnetic reflections both at
the Mn $K$ edge and the Eu absorption edges when it comes to the
intensity- and $\vecc{q}$-dependence with varying temperature and
photon polarization, along with an order parameter-like behaviour of
the Eu magnetic order.  Hence the Eu resonant reflections are caused
by magnetic order of the Eu$^{3+}$ \sfo\ moments which mirrors the
magnetic structure of Mn moments.  The \sfo\ state of Eu$^{3+}$ can be
induced by several symmetry breaking mechanisms, of which two present
themselves as the most likely.  The fact that the Eu moments mirror
the Mn magnetic structure clearly shows the presence of an exchange
field at the Eu sites.  This exchange field will cause a local
symmetry breaking at the Eu sites that, like the symmetry breaking by
an external magnetic field, can result in a population of the \sfo\
state. In a similar way, a local symmetry breaking due to the crystal
field could also populate the \sfo\ excited state, i.e. an electric
field could transform the non-magnetic ion into a magnetic one. 
Whether the symmetry is broken primarily by the crystal
field or the Mn exchange field, the populated \sfo\ state at the Eu
sites is exchange coupled to the Mn magnetic sublattice. In principle,
this scenario also allows for Eu-Eu exchange
interaction~\cite{Palermo1980}.  The mechanism here is very different
from the origin of the magnetic polarization observed at oxygen sites
in similar multiferroics, which is caused by spin dependent
hybridization~\cite{Bea10,Par11}.

The existence of a non-zero Eu magnetic moment in \euya\ raises a
couple of issues.  Firstly, it puts into question the assumption of
\euyx\ being a model system for multiferroic orthomanganites free from
RE magnetism.  As soon as the magnetic Eu state is induced, Eu-Mn and
Eu-Eu exchange coupling as well as Eu magnetic anisotropy can
contribute to the magnetic order of the entire system with the
possibility of an indirect impact on the field-dependent
multiferroicity.  Furthermore, similar RE magnetic structures have
been shown to contribute directly to ferroelectricity in \rmo\
(\textit{R} = Tb, Dy, Gd) by symmetric exchange striction
mechanisms~\cite{Kat05,Ser06,Mos06,Ali08,Fey09,Fey10,Wal11,Ska14}.

Secondly, the discovery of a Eu magnetic moment without self-ordering
opens up the possibility of using Eu as a magnetic probe.  As seen in
Fig.~\ref{fig:tdep}, by making use of the strong resonant enhancement
at the Eu $M_{4,5}$ edges we gain a drastically stronger signal from
which we are able to extract intensities and $q$-values to much higher
precision than what is possible at the Mn $K$ edge. The onset of the
ferroelectric phase below $\sim$ 28 K is clearly seen.  Measuring at
the Eu $M_{4,5}$ edges also has the added benefit of being able to
explore a larger Ewald sphere compared to the Mn $L_{2,3}$ edges,
where detailed studies of Mn magnetism are usually performed.

Thirdly, our findings may have implications for magnetism in other
transition metals.  Van Vleck effects are expected to appear in Eu,
but not in $3d$ transition metals, where the orbital moment is usually
quenched.  There are however Van Vleck materials where exchange
coupling plays a role and antiferromagnetic order in a field is
associated with Bose-Einstein condensation of magnons~\cite{Zap06}.
More significantly, the increasing spin-orbit interaction in $4d$ and
$5d$ transition metals renders Van Vleck effects more important in
this class of materials that are increasingly attracting
interest~\cite{Kha13}.

In conclusion, we have presented the first observation of long range
antiferromagnetic order of Van Vleck ions, where the ordering is due
to exchange coupling between Eu and Mn spins.  Since an exchange
interaction between Eu$^{3+}$ $4f$ moments could already, in
principle, intermix \sfo\ contributions, our observation allows for
the hope to observe complex long range magnetic order in systems
consisting entirely of formally non-magnetic ions, where novel
properties may be expected.

\end{document}